\documentclass[floatfix,amsfonts,amssymb,amsmath,superscriptaddress,a4paper,showkeys,aps,prd,11pt]{revtex4-2}

\usepackage[utf8]{inputenc}
\usepackage[T1]{fontenc}
\usepackage{array}
\usepackage{xcolor}
\usepackage{graphicx}
\usepackage{caption}
\usepackage{subcaption}
\usepackage{tabularx}
\usepackage{booktabs}
\usepackage{hyperref}
\hypersetup{colorlinks=true, citecolor=blue, urlcolor=blue, linkcolor=blue}

\begin{document}

\title{Search for R-Parity-Violation-Induced Charged Lepton Flavor Violation at Future Lepton Colliders}

\author{Xunye Cai}
\author{Jingshu Li}
\author{Ran Ding}
\author{Meng Lu}
\author{Zhengyun You}
\email[]{youzhy5@mail.sysu.edu.cn}
\affiliation{
School of Physics, Sun Yat-Sen University, Guangzhou 510275, China}

\author{Qiang Li}
\email[]{qliphy0@pku.edu.cn}
\affiliation{
School of Physics and State Key Laboratory of Nuclear Physics and Technology, Peking University, Beijing, 100871, China}

\begin{abstract}
Interest in searches for Charged Lepton Flavor Violation~(CLFV) has continued in the past few decades since the observation of CLFV would indicate a new physics Beyond the Standard Model~(BSM). As several future lepton colliders with high luminosity have been proposed, the search for CLFV will reach an unprecedented level of precision. Many BSM models allow CLFV processes at the tree level, such as the R-parity-violating~(RPV) Minimal Supersymmetric Standard Model~(MSSM), which is a good choice for benchmarking. In this paper, we perform a detailed fast Monte Carlo simulation study on RPV-induced CLFV processes at future lepton colliders, including a $240$ $\mathrm{GeV}$ circular electron positron
collider~(CEPC) and a $6$ or $14$ $\mathrm{TeV}$ Muon Collider. As a result, we found that the upper limits on the $\tau$-related RPV couplings will be significantly improved, while several new limits on RPV couplings can be set, which are inaccessible by low-energy experiments.
\end{abstract}


\maketitle
\tableofcontents


\section{Introduction}
\label{sec:intro}

Although the Standard Model~(SM) has achieved great success in the field of particle physics, it is still an incomplete theory. In the SM, lepton numbers are global $U(1)$ symmetries;~and thus the electron lepton number $L_e$, muon lepton number $L_\mu$ , and tau lepton numbers $L_\tau$ are separately conserved, as is the total lepton number $L=L_e+L_\mu+L_\tau$. However, it is not consistent with the discovery of neutrino oscillations and non-zero neutrino masses, demonstrating that these symmetries are accidental and that there could be a lepton-flavor-violating short-range interaction among the charged leptons~\cite{Gonzalez-Garcia:2002bkq}. 
Therefore, Charged Lepton Flavor Violation~(CLFV) processes~\cite{deGouvea:2013zba,Davidson:2022jai,Bernstein:2013hba} are expected to occur. However, even in the SM extended with a non-zero mass neutrino, CLFV rates are typically suppressed by a factor of $G_F^2 m_v^4 \sim 10^{-50}$, which is well below the sensitivity of the experiment and should be unobservable. 

Nevertheless, many models of physics Beyond the Standard Model~(BSM) introduce new sources of CLFV, such as Supersymmetry (SUSY)~\cite{Lee:1984kr,Lee:1984tn}, $Z^\prime$ boson~\cite{Langacker:2008yv}, 
leptoquark~\cite{Dorsner:2016wpm}, quantum black hole~(QBH) in low-scale gravity~\cite{Gingrich:2009hj}, two-Higgs-doublet model~(2HDM)~\cite{Branco:2011iw}, and R-parity-violating~(RPV) Minimal Supersymmetric Standard Model~(MSSM)~\cite{Choudhury:2024ggy,Chemtob:2004xr}. These models can give rise to sizeable CLFV rates that may be detectable in the next generation of collider experiments. Any such detection of CLFV would be clear evidence for the existence of New Physics~(NP) and shed light on the probe of BSM physics. 

For the reasons above, the search for CLFV has attracted great interest in recent decades. Numerous experiments have been performed or will be constructed with different approaches, including muon-based experiments, such as $\mu^{-} N \rightarrow e^{-} N$ at Mu2e~\cite{Mu2e:2014fns,Mu2e:2022ggl} and COMET~\cite{COMET:2018auw}, $\mu \rightarrow e \gamma$ at MEG-II~\cite{Renga:2022tex,MEGII:2021fah}, $\mu \rightarrow e e e$ at Mu3e~\cite{Mu3e:2020gyw,Dittmeier:2022nwi}, and  high energy colliders like LEP and LHC looking for CLFV decays of mesons~\cite{CLEO:2008lxu,BaBar:2021loj,BESIII:2022exh,Achasov:2009en}, $\mu$, $\tau$~\cite{BaBar:2009hkt}, $Z$~bosons~\cite{ATLAS:2014vur,OPAL:1995grn,DELPHI:1996iox}, and Higgs bosons~\cite{ATLAS:2019pmk,CMS:2017con,ATLAS:2019old}. In the next decades, several proposed lepton colliders, such as the Circular Electron Positron Collider~(CEPC)~\cite{CEPCStudyGroup:2018rmc},  the Future Circular Collider~(FCC)-ee~\cite{Fccweb},  the International Linear Collider~(ILC)~\cite{Barklow:2015tja}, the  Compact Linear Collider~(CLIC)~\cite{CLICdp:2018cto} and  the Muon Collider~\cite{Delahaye:2019omf}, will be ideal facilities to further probe   CLFV processes~\cite{Li:2018cod,Li:2021lnz,Homiller:2022iax,Bossi:2020yne,Cirigliano:2021img} since they have cleaner environments with low backgrounds and higher luminosity than hadron colliders.

In a previous work~\cite{Li:2023lin}, the potential to search for CLFV signals induced by the Z$^{\prime}$ model has been studied at the future lepton colliders.
In this paper, we focus on the search for CLFV at CEPC and the Muon Collider with RPV-MSSM assumed. 
The Z$^{\prime}$ model assumes the existence of an extra Z$^{\prime}$ boson that couples to different lepton flavors, while RPV-MSSM is another interesting new physics model based on SUSY.
The rest of this paper is organized as follows. In Section~\ref{sec:RPV}, we will give a brief introduction to RPV-MSSM and its present research status. Section~\ref{sec:detail} discusses the details of the fast Monte Carlo simulation we performed at CEPC and Muon Collider. In Section~\ref{sec:result}, we present the numerical results of RPV coupling limits and compare them with current and prospective experimental limits from low-energy $\mu$ and $\tau$ experiments. Lastly, we close with a conclusion of this paper in Section~\ref{sec:conclusion}.


\section{R-Parity Violating MSSM}
\label{sec:RPV}

MSSM is one of the promising candidates for BSM physics. In the MSSM, renormalizability and gauge invariance do not forbid all the coupling terms that cause lepton number and baryon number violation. It can be prevented by introducing a $\mathbb{Z}_2$ symmetry called R-parity~\cite{Barbier:2004ez}.
\begin{equation}
R_p := (-1)^{3 B+L+2 S},
\end{equation} 
where $B$, $L$, and $S$ denote the baryon number, lepton number, and spin of the particle, respectively. All the SM particles have an R-parity of $+1$, while all the SUSY particles have an R-parity of $-1$. One of the motivations for this symmetry is to ensure the stability of the lightest supersymmetric particle, which is a possible dark matter candidate~\cite{KumarBarman:2020ylm,Barman:2022jdg,Chowdhury:2016qnz}. R-parity conservation in MSSM will result in a large transverse missing energy signature at the collider experiment~\cite{ATLAS:2022hbt,ATLAS:2022ckd,ATLAS:2022zwa,CMS:2021few,CMS:2022vpy,CMS:2022sfi}, but no such signals have been observed so far. Furthermore, RPV-MSSM can give a substantial contribution to muon $g-2$ calculation through bilinear and trilinear terms in the RPV superpotential~\cite{Martin:2001st,Moroi:1995yh,Chakraborty:2015bsk}. Allowing R-parity to be broken becomes acceptable and will give rise to a series of phenomenological consequences; thus, R-parity violation has been extensively studied in phenomenological theory~\cite{Kohler:2024mpu,Calibbi:2021fld,Hu:2018lmk,Bardhan:2021adp,Zheng:2022ssr,Zheng:2022irz,Zheng:2021wnu,Hu:2020yvs} and in collider experiments~\cite{Karmakar:2023mhr,ATLAS:2021byo,Dercks:2017lfq,Yamanaka:2014oia}.

When R-parity is broken, the  R-parity violating superpotential must be included:

\begin{equation}
W_{R P V}=\frac{1}{2} \lambda_{i j k} L_i L_j E_k^c+\lambda_{i j k}^{\prime} L_i Q_j D_k^c+\frac{1}{2} \lambda_{i j k}^{\prime \prime} U_i^c D_j^c D_k^c+\mu_i L_i H_u,
\end{equation}
where $L$,$E$,$Q$,$U$, and $D$ are superfields of lepton, charged lepton, quark, up quark, and down quark respectively; $H_u$ is one of the Higgs superfields; $\lambda_{i j k}$, $\lambda_{i j k}^\prime$, $\lambda_{i j k}^{\prime \prime}$ are Yukawa couplings; and $i,j,k$ denote the three generations. Gauge invariance enforces the antisymmetry of two indices for these couplings, $\lambda_{i j k}=-\lambda_{j i k}$ and $\lambda_{i j k}^{\prime \prime}=-\lambda_{i k j}^{\prime \prime}$. We can write down the RPV part of the interaction Lagrangian in terms of component fields.

\begin{equation}
\begin{aligned}
\mathcal{L}_{I}= & -\frac{1}{2} \lambda_{i j k}\left(\tilde{\nu}_{i L} \bar{l}_{k R} l_{j L}+\tilde{l}_{j L} \bar{l}_{k R} \nu_{i L}+\tilde{l}_{k R}^* \bar{\nu}_{i R}^c l_{j L}-(i \leftrightarrow j)\right) \\
& -\lambda_{i j k}^{\prime}\left(\tilde{\nu}_{i L} \bar{d}_{k R} d_{j L}+\tilde{d}_{j L} \bar{d}_{k R} \nu_{i L}+\tilde{d}_{k R}^* \bar{\nu}_{i R}^c d_{j L}\right. \\
& \left.-\tilde{l}_{i L} \bar{d}_{k R} u_{j L}-\tilde{u}_{j L} \bar{d}_{k R} l_{i L}-\tilde{d}_{k R}^* \bar{l}_{i R}^c u_{j L}\right)+ \text { h.c.},
\end{aligned}
\end{equation}
where $\lambda^{\prime\prime}=0$ is assumed for simplicity since non-zero $\lambda^{\prime\prime}$ corresponds to baryon number violation, which is irrelevant to the search for CLFV. $l,\nu,u,d$ denote the field of charged lepton, neutrino, up quark, and down quark, respectively. The tilde over a field represents its superpartner field. The superscripts ${}^c$ and ${}^*$ represent charge conjugation and complex conjugation. The subscripts $L$ and $R$ represent left-handed and right-handed fields.

The interaction Lagrangian can allow for several CLFV processes at the tree level. In this work, we focus on CLFV processes produced by the above $\lambda_{i j k}\tilde{\nu}_{i L} \bar{l}_{k R} l_{j L}$ term, which can contribute to lepton collision 
CLFV processes at lepton colliders. There have been some studies similar to this paper performed to search for CLFV based on RPV-MSSM, but with different CLFV processes like lepton and meson decay and were studied at different experimental facilities like LHC and COMET~\cite{ATLAS:2018mrn,Sato:2014ita,Dreiner:2006gu,Choudhury:1996ia,Huitu:1997bi,Kim:1997rr,Dreiner:2001kc,Littenberg:2000fg}.

\section{Simulation and Analysis Framework}
\label{sec:detail}
In this manuscript, we focus on the CLFV search based on RPV-MSSM and perform the simulation at a $240$ $\mathrm{GeV}$ electron--positron collider, i.e., the CEPC with an integrated luminosity of $5$ $\mathrm{ab}^{-1}$ and a $6$ or $14$ $\mathrm{TeV}$ Muon Collider with an integrated luminosity of~$4$~$\mathrm{ab}^{-1}$.

\subsection{Event Simulation}

\begin{figure}[ht]
  \centering
  \begin{subfigure}[htbp]{0.49\textwidth}
    \centering
    \includegraphics[width=0.9\textwidth]{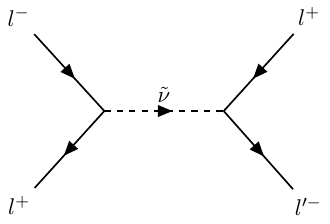}    
    \caption{\(s\) channel diagram}
  \end{subfigure}
  \begin{subfigure}[htbp]{0.49\textwidth}
    \centering
    \includegraphics[width=0.6\textwidth]{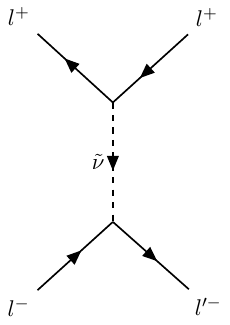} 
    \caption{\(t\) channel diagram}
  \end{subfigure}
  \caption{Feynman diagrams of CLFV signal processes, namely the  s-channel process (a) and the t-channel process (b), both propagated by s-neutrino.}
  \label{FeynDiagram}
\end{figure}

The CLFV signal processes studied in this manuscript include $ee\rightarrow e\mu$, $ee\rightarrow e\tau$, $ee\rightarrow \mu\tau$ at $240$ $\mathrm{GeV}$ CEPC and $\mu\mu\rightarrow e\mu$, $\mu\mu\rightarrow e\tau$, $\mu\mu\rightarrow \mu\tau$ at 6 or 14 TeV Muon Collider. 
Figure~\ref{FeynDiagram} gives some examples of Feynman diagrams for these CLFV signal processes. The main background processes for each signal process are summarized in Table~\ref{BkgTable}, where the $\mathrm{WW}$ and $\tau\tau$ background mean $\ell \ell \rightarrow \mathrm{WW} \text{ or }\tau\tau$, with both $\mathrm{W}$ or $\tau$ decaying into the corresponding charged leptons in the final state. For the $\tau$-related signal channel like $ee\rightarrow e\tau$, the $\tau\tau$ background process has only one $\tau$ decaying into the charged lepton, while the other $\tau$ goes through hadronic decay and is reconstructed with the jet collection in $\textsc{Delphes}$ and likewise for the $\mathrm{H} \nu \bar{\nu}(\mathrm{H} \rightarrow \tau \tau)$ and $\mathrm{H} \nu \bar{\nu}(\mathrm{H} \rightarrow \mathrm{WW})$ background processes.

\begin{table}[ht] 
\caption{Summary of the CLFV signal and background processes.\label{BkgTable}}
\newcolumntype{C}{>{\centering\arraybackslash}X}
\begin{tabularx}{\textwidth}{cC}
\toprule
\textbf{Signal Process} & \textbf{ Background Processes}\\
\midrule$e e \rightarrow e \mu$ & $\mathrm{WW}, \mathrm{H} \nu \bar{\nu}(\mathrm{H} \rightarrow \tau \tau), \mathrm{H} \nu \bar{\nu}(\mathrm{H} \rightarrow \mathrm{WW}), \tau \tau$ \\
$e e \rightarrow e \tau$ & $\mathrm{WW}, \mathrm{H} \nu \bar{\nu}(\mathrm{H} \rightarrow \tau \tau), \tau \tau$ \\
$e e \rightarrow \mu \tau$ & $\mathrm{WW}, \mathrm{H} \nu \bar{\nu}(\mathrm{H} \rightarrow \tau \tau), \tau \tau$ \\
$\mu \mu \rightarrow e \mu$ & $\mathrm{WW}, \mathrm{WW} \nu \bar{\nu}, \mathrm{H} \nu \bar{\nu}(\mathrm{H} \rightarrow \tau \tau), \mathrm{H} \nu \bar{\nu}(\mathrm{H} \rightarrow \mathrm{WW}), \tau \tau$ \\
$\mu \mu \rightarrow e \tau$ & $\mathrm{WW}, \mathrm{WW} \nu \bar{\nu}, \mathrm{H} \nu \bar{\nu}(\mathrm{H} \rightarrow \tau \tau), \mathrm{H} \nu \bar{\nu}(\mathrm{H} \rightarrow \mathrm{WW}), \tau \tau$ \\
$\mu \mu \rightarrow \mu \tau$ & $\mathrm{WW}, \mathrm{WW} \nu \bar{\nu}, \mathrm{H} \nu \bar{\nu}(\mathrm{H} \rightarrow \tau \tau), \mathrm{H} \nu \bar{\nu}(\mathrm{H} \rightarrow \mathrm{WW}), \tau \tau$ \\
\bottomrule
\end{tabularx}
\end{table}

Using the UFO~\cite{Degrande:2011ua} model published on the FeynRules model database for RPV extension of MSSM~\cite{Fuks:2012im}, both signal and background process events are generated with MadGraph5\_aMC@NLO version 3.4.2~\cite{Alwall:2011uj,Alwall:2014hca}, and then we perform parton shower and hadronization with $\textsc{Pythia8}$ version 3.0.6~\cite{Sjostrand:2014zea}. In particular, the initial-state radiation (ISR) effect~\cite{Frixione:2021zdp} was incorporated into the simulation. Lastly, we used $\textsc{Delphes}$ version 3.5.0~\cite{deFavereau:2013fsa} for detector fast simulation with the default detector configuration cards of CEPC and the Muon Collider.

\subsection{Event Selection and Analysis Method}
\label{sec:method}

The event selection criteria are described as follows. First, the events must include exactly two charged leptons in the final state with transverse momentum $p_T$ and pseudo-rapidity $\eta$ satisfying $p_T > 10$ GeV/$c$ and $|\eta| < 2.5$. In particular, for the $\tau$-related channel, $\tau$ goes through hadronic decay and is reconstructed with the jet collection in $\textsc{Delphes}$, and the jets must satisfy $p_T > 20$ GeV/$c$ and $|\eta| < 5$. These basic cuts applied on $p_T$ and $|\eta|$ follow the default detector configuration cards of $\textsc{Delphes}$, reflecting the tracking information provided by the detector. In addition, the final state leptons must meet the requirements of lepton flavor change and charge conservation. For example, in the \mbox{$e^{+} e^{-} \to e^{-} \mu^{+} ( e^{+} \mu^{-})$ }channel, all events must have only one $e^{-} (e^{+})$ and one $\mu^{+} (\mu^{-})$.

For CEPC, the $\mu$ tracking efficiency $\epsilon$ is set to be 100\% within $0.1<|\eta|\leqslant 3$, and 0\% for $|\eta|>3 \text{ or } |\eta|\leqslant 0.1$. For the Muon Collider, the $\mu$ tracking efficiency is $\epsilon\geqslant90\%$ within $|\eta|\leqslant 2.5$, and 0\% for $|\eta| > 2.5$. For the $\tau$-related channel, as defined in $\textsc{Delphes}$ default cards of the detector configuration, the $\tau$ tagging efficiency is assumed to be 40\% for the CEPC and 80\% for the Muon Collider with $p_T > 10$ GeV/$c$.

\begin{figure}[ht]
  \centering
  \begin{subfigure}[htbp]{0.49\textwidth}
    \centering
    \includegraphics[width=\textwidth]{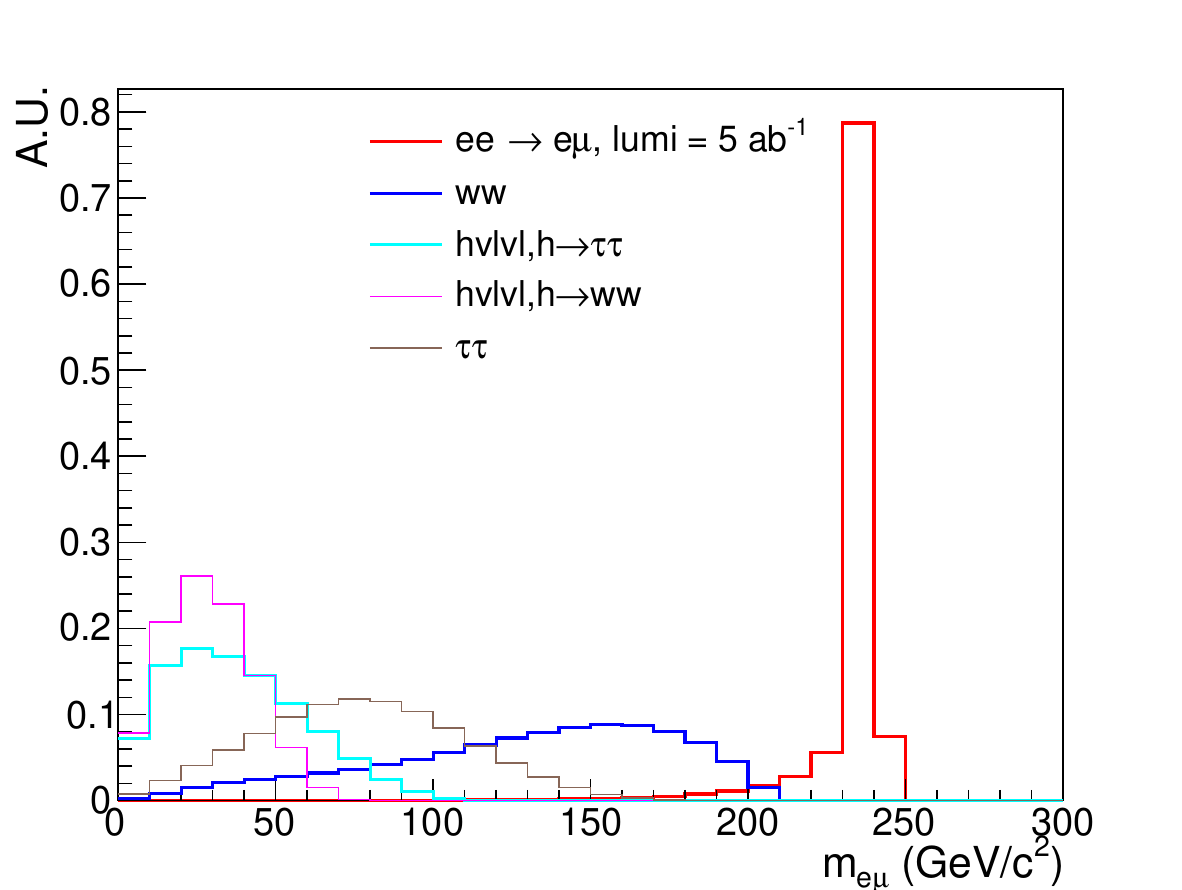}
    \caption{}
    \label{eeemumll}
  \end{subfigure}
  \hfill
  \begin{subfigure}[htbp]{0.49\textwidth}
    \centering
    \includegraphics[width=\textwidth]{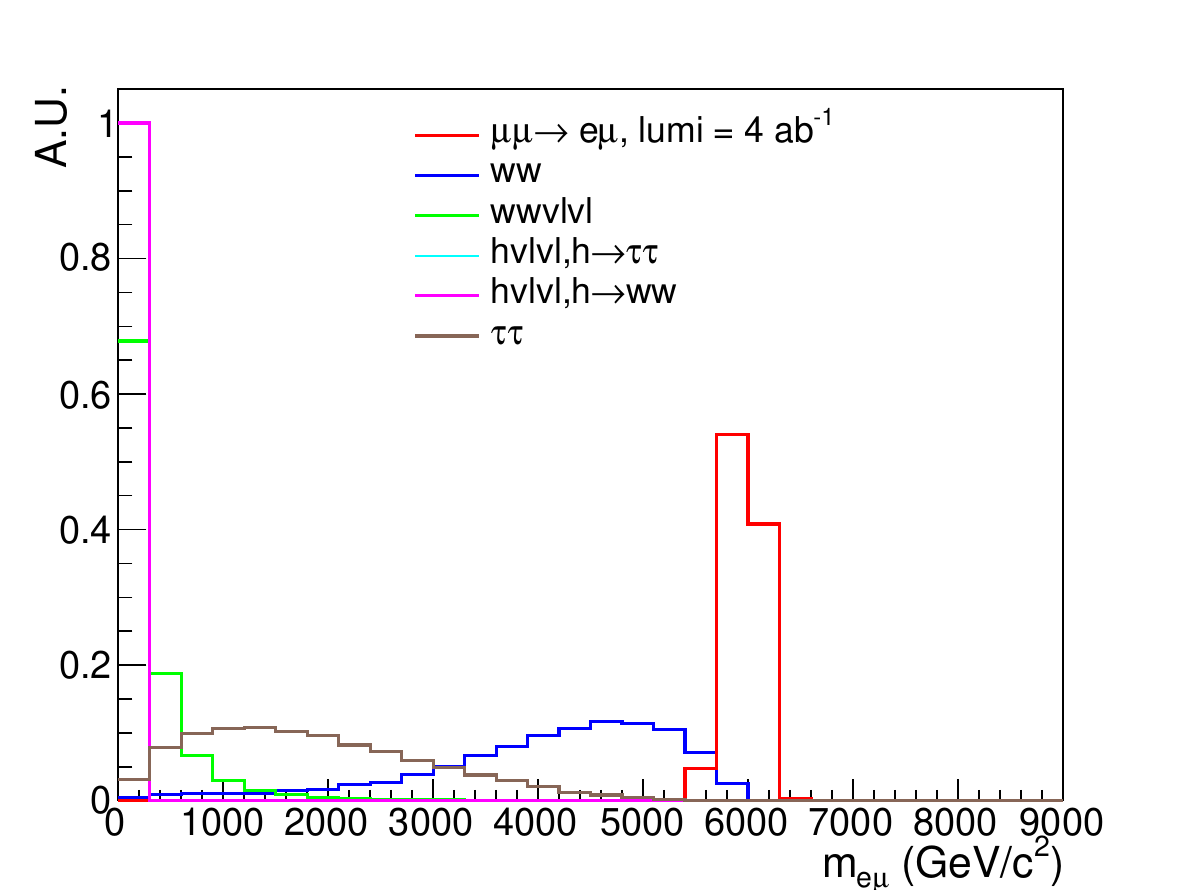}
    \caption{}
    \label{mumuemumll}
  \end{subfigure}
  \newline
  \begin{subfigure}[htbp]{0.49\textwidth}
    \centering
    \includegraphics[width=\textwidth]{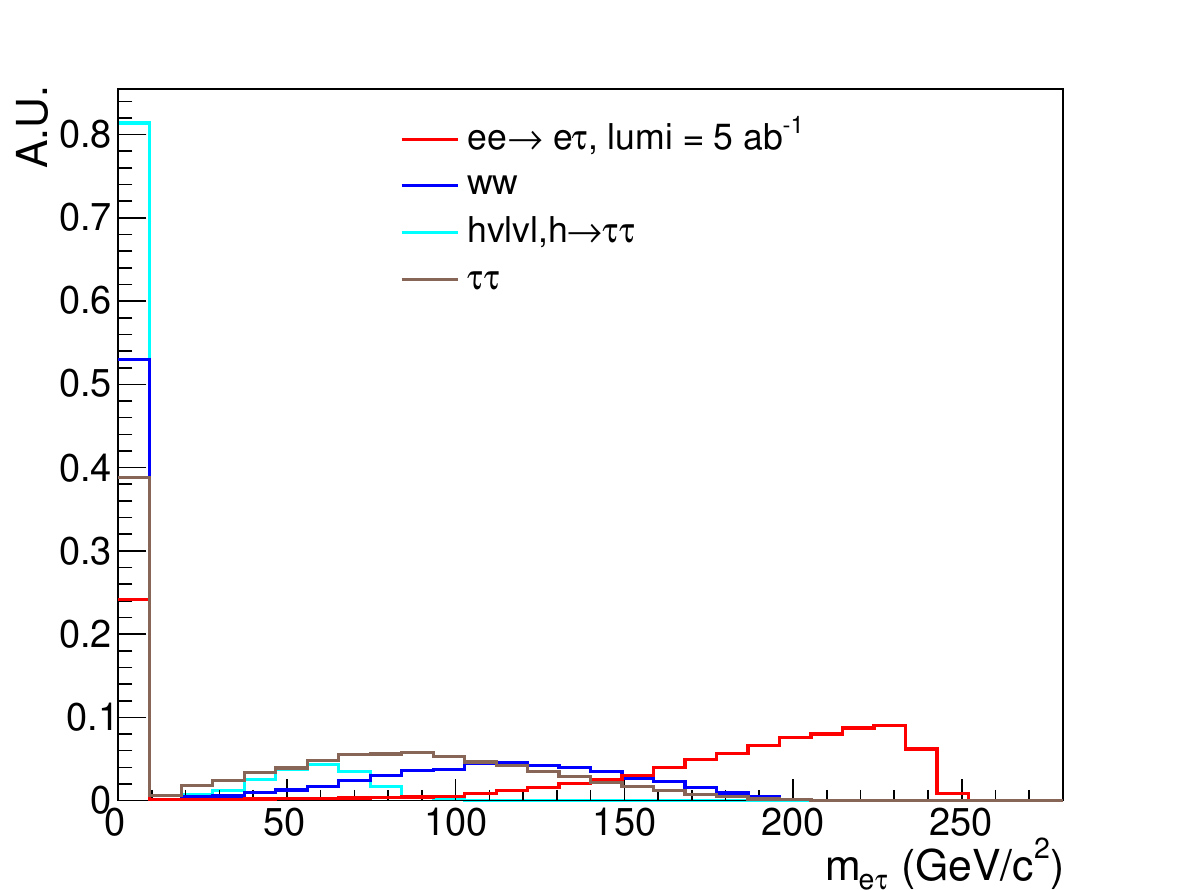}
    \caption{}
    \label{eeetaumll}
  \end{subfigure}
  \hfill
  \begin{subfigure}[htbp]{0.49\textwidth}
    \centering
    \includegraphics[width=\textwidth]{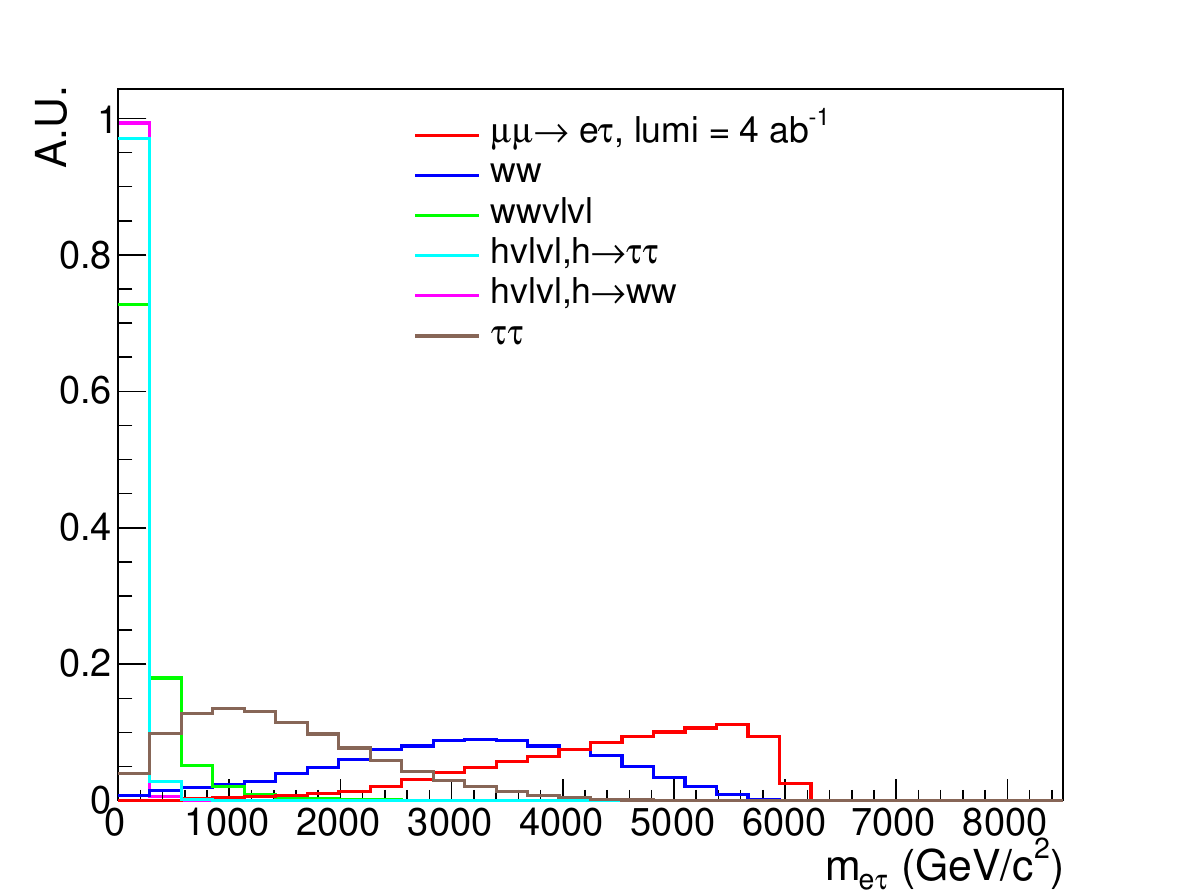}
    \caption{}
    \label{mumuetaumll}
  \end{subfigure}
  \newline
  \begin{subfigure}[htbp]{0.49\textwidth}
    \centering
    \includegraphics[width=\textwidth]{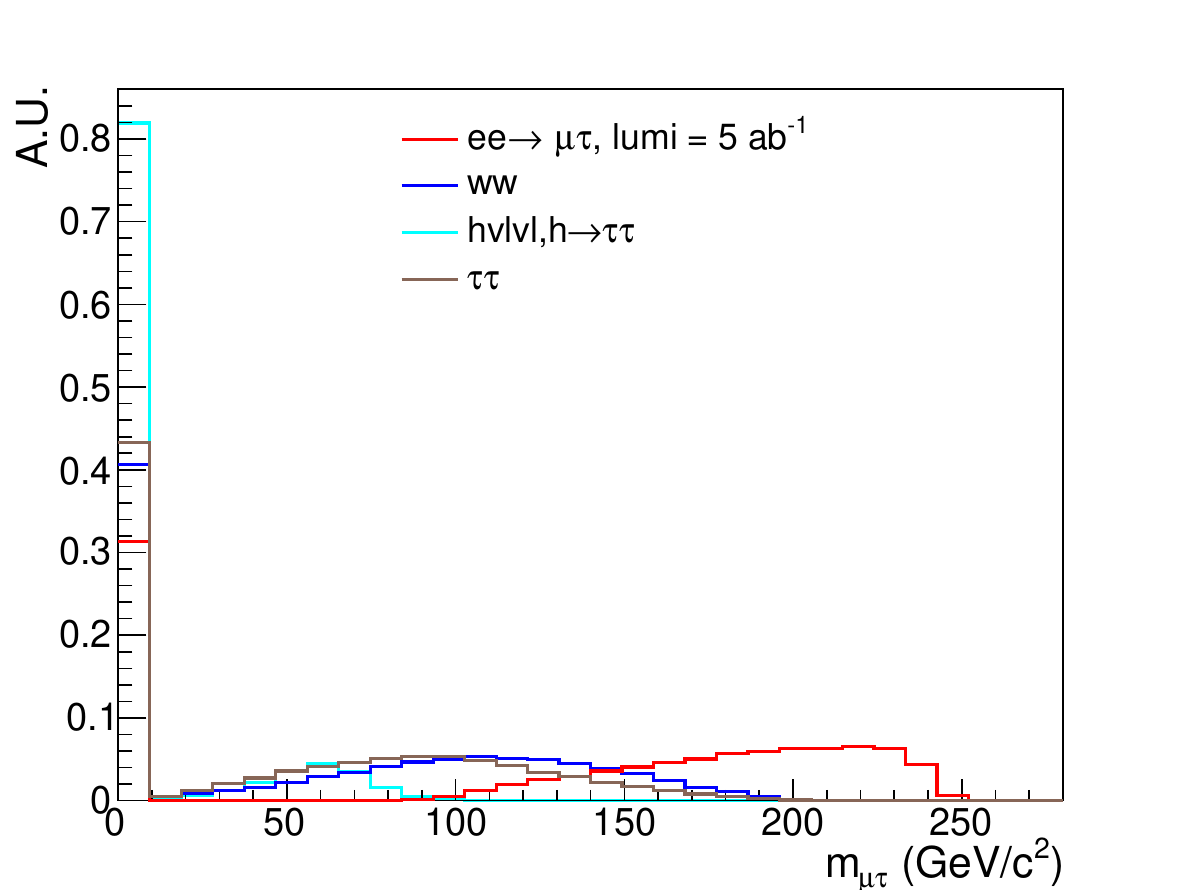}
    \caption{}
    \label{eemutaumll}
  \end{subfigure}
  \hfill
  \begin{subfigure}[htbp]{0.49\textwidth}
    \centering
    \includegraphics[width=\textwidth]{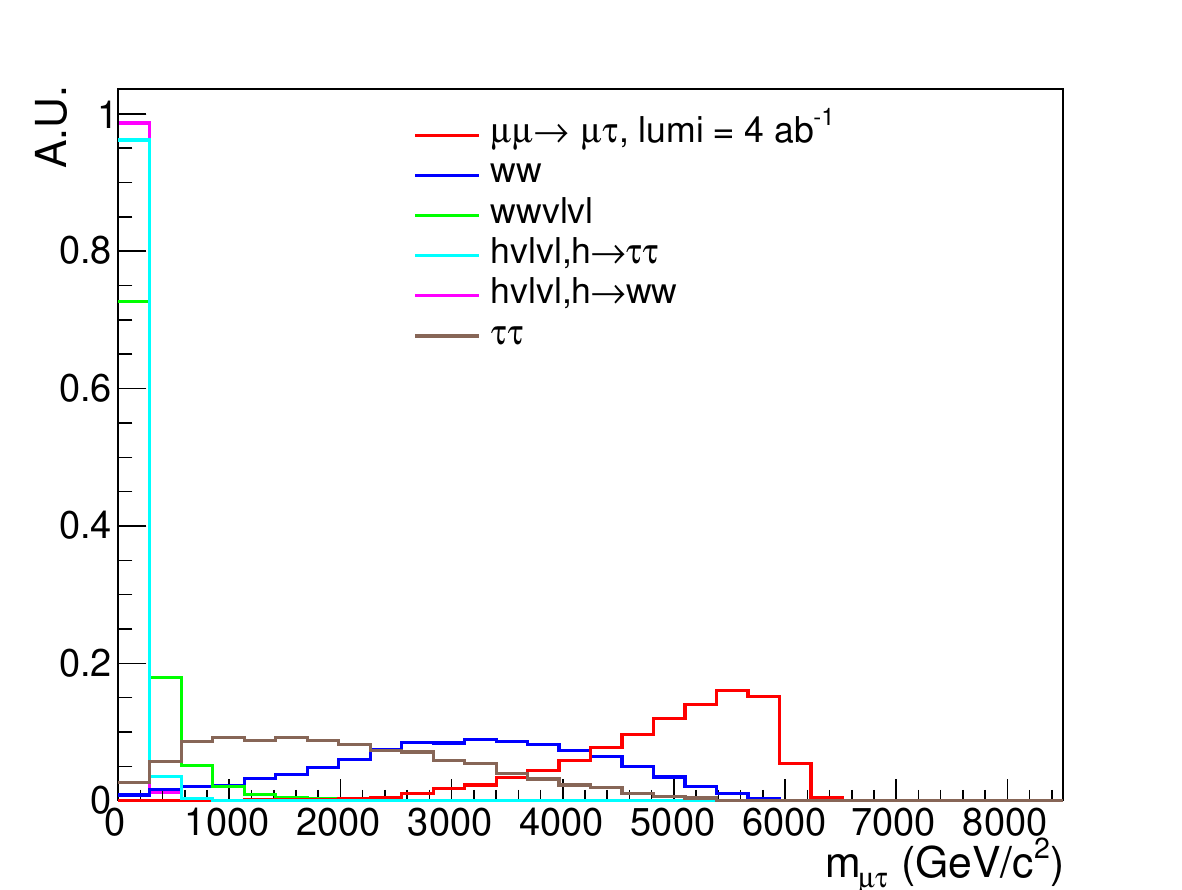}
    \caption{}
    \label{mumumutaumll}
  \end{subfigure}
  \caption{Invariant mass distributions of the final state di-leptons for $ee\rightarrow e\mu$ (a), $ee\rightarrow e\tau$ (c), and $ee\rightarrow \mu\tau$ (e) channel at CEPC, and $\mu\mu\rightarrow e\mu$ (b), $\mu\mu\rightarrow e\tau$ (d), and $\mu\mu\rightarrow \mu\tau$ (f) channel at the $6$ $\mathrm{TeV}$ Muon Collider. A.U. refers to the Arbitrary Unit.}
  \label{massfig}
\end{figure}

\begin{figure}[ht]
  \centering
  \begin{subfigure}[htbp]{0.49\textwidth}
    \centering
    \includegraphics[width=\textwidth]{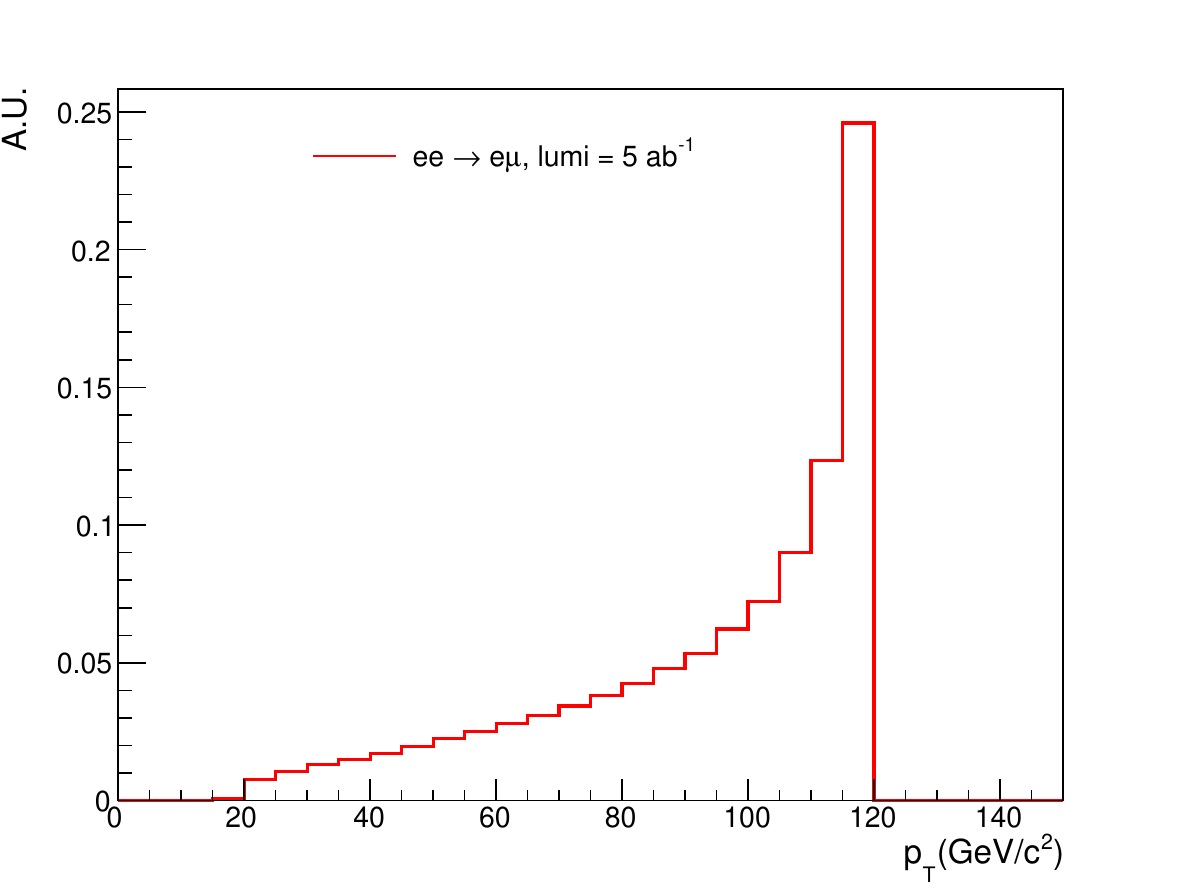}
    \caption{}
  \end{subfigure}
  \hfill
  \begin{subfigure}[htbp]{0.49\textwidth}
    \centering
    \includegraphics[width=\textwidth]{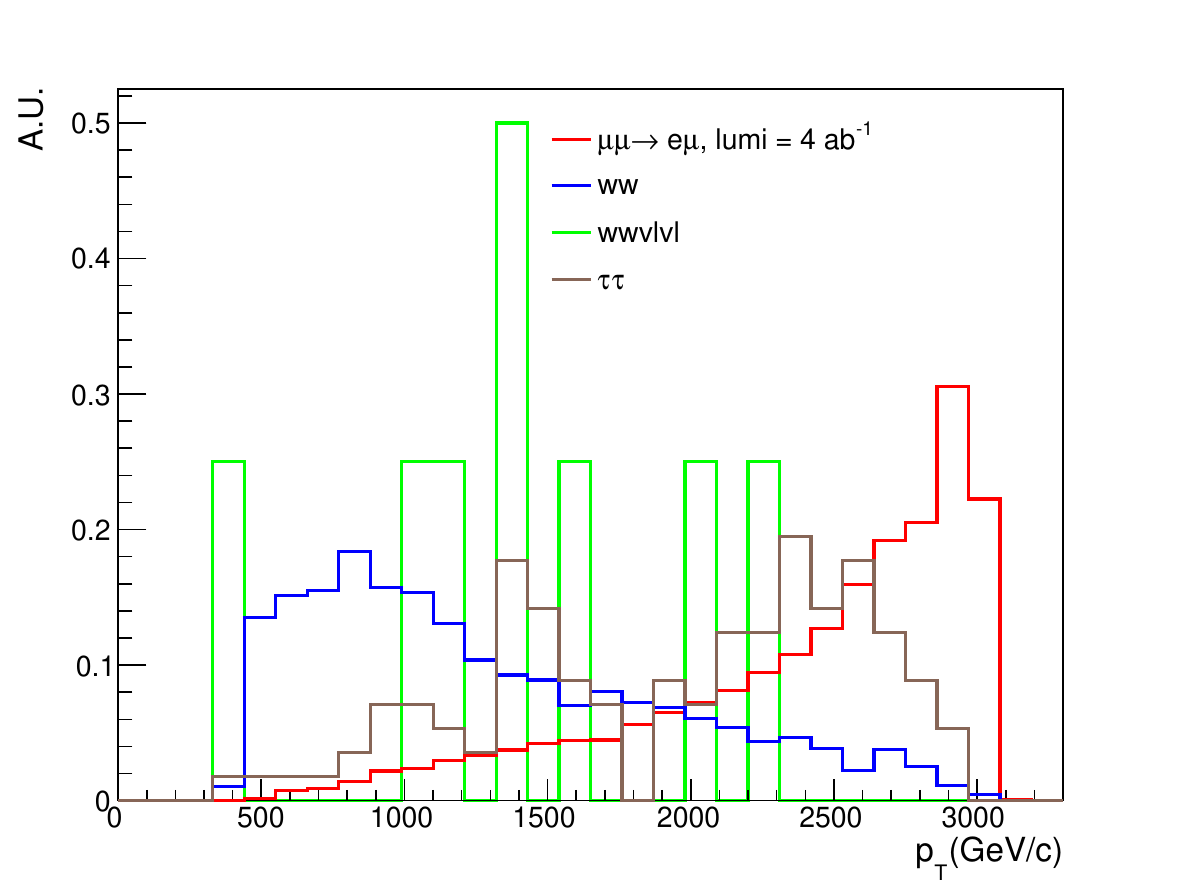}
    \caption{}
  \end{subfigure}
  \newline
  \begin{subfigure}[htbp]{0.49\textwidth}
    \centering
    \includegraphics[width=\textwidth]{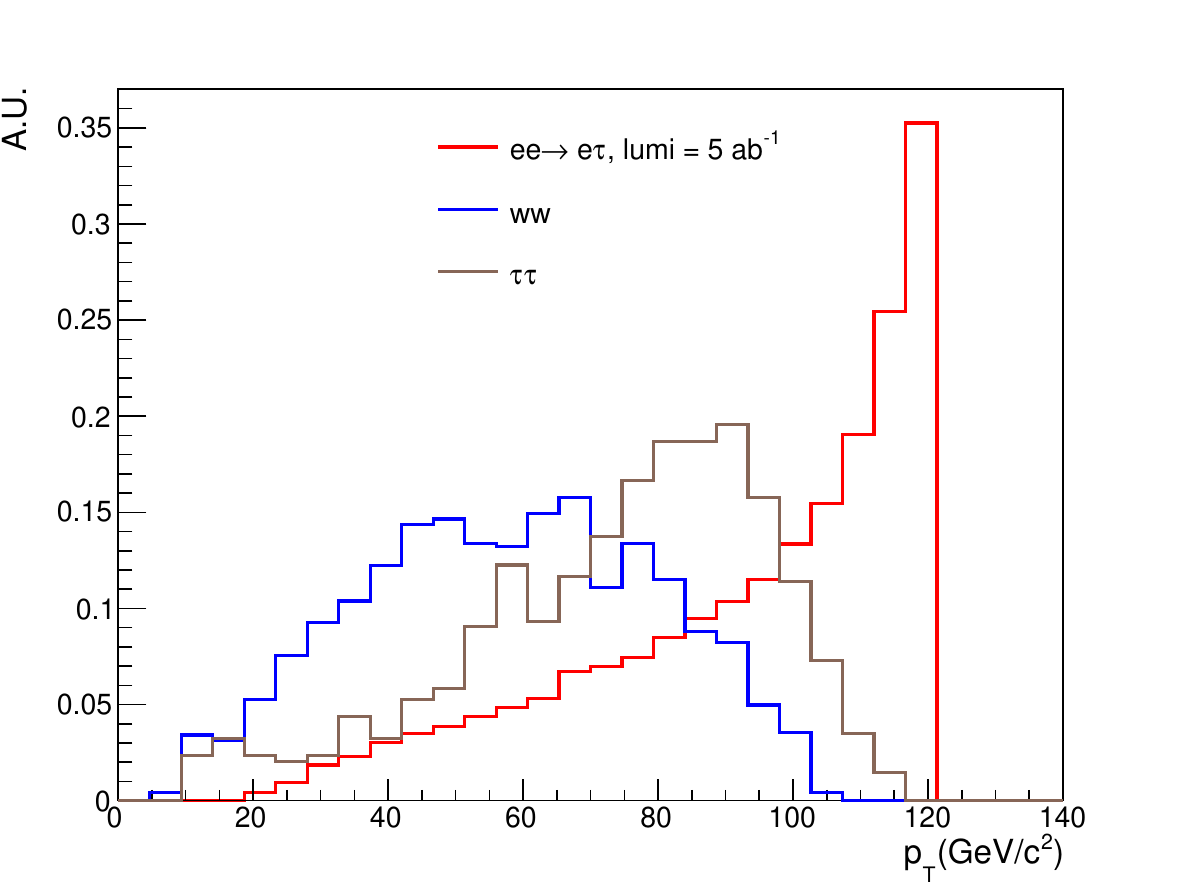}
    \caption{}
  \end{subfigure}
  \hfill
  \begin{subfigure}[htbp]{0.49\textwidth}
    \centering
    \includegraphics[width=\textwidth]{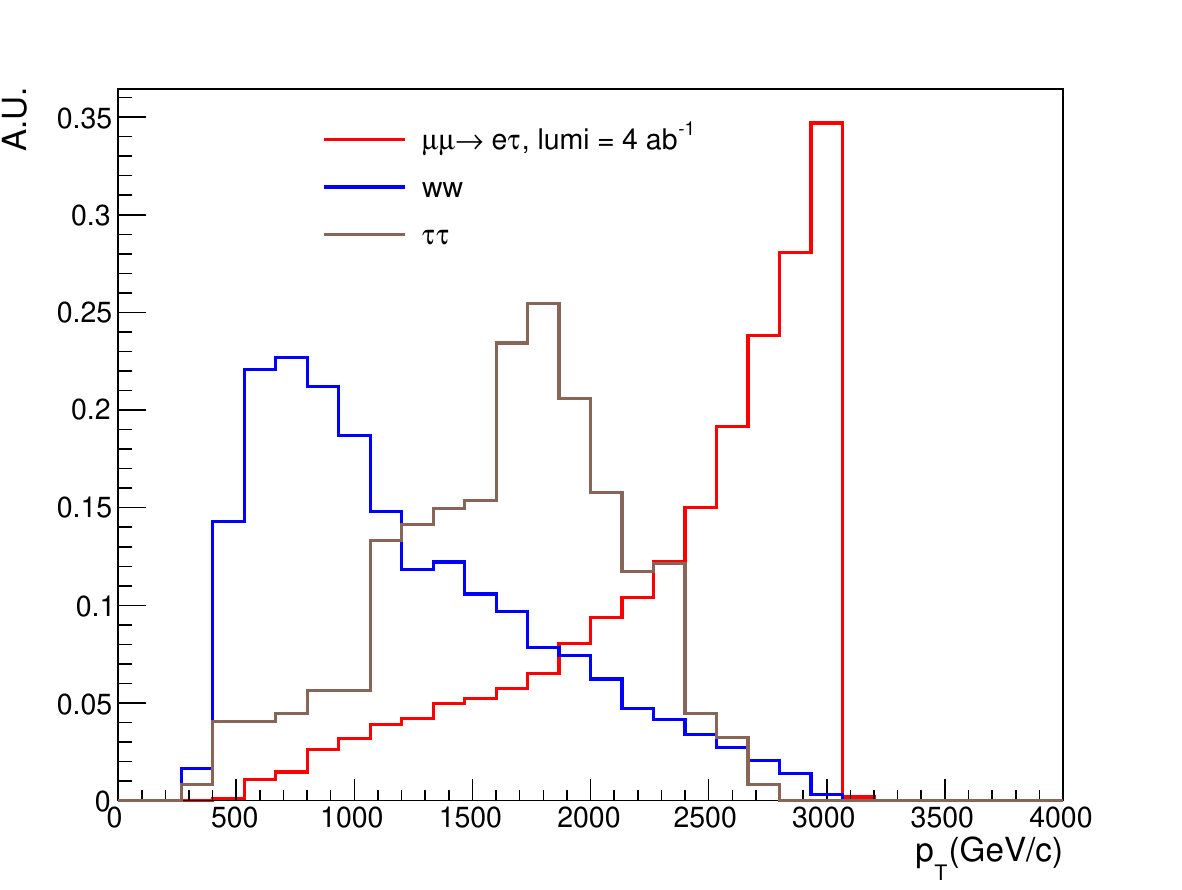}
    \caption{}
  \end{subfigure}
  \newline
  \begin{subfigure}[htbp]{0.49\textwidth}
    \centering
    \includegraphics[width=\textwidth]{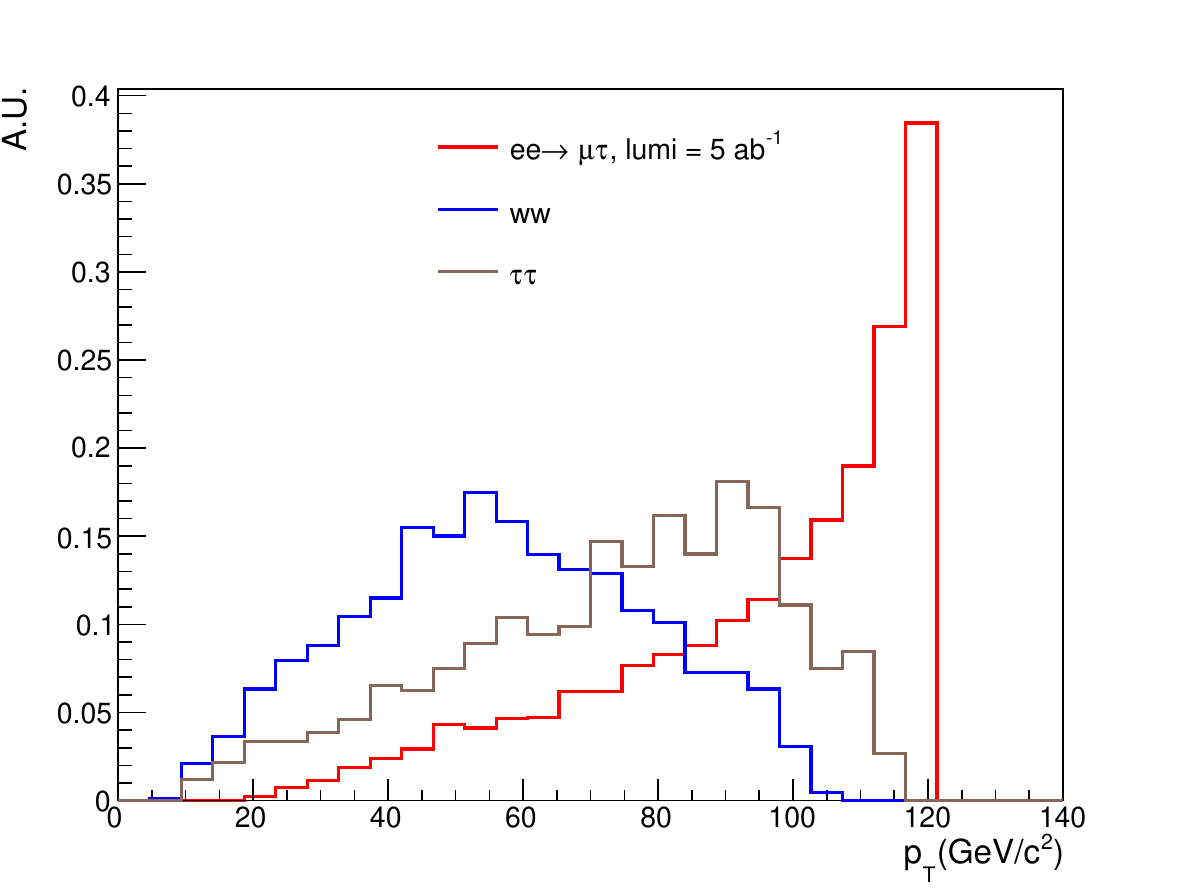}
    \caption{}
  \end{subfigure}
  \hfill
  \begin{subfigure}[htbp]{0.49\textwidth}
    \centering
    \includegraphics[width=\textwidth]{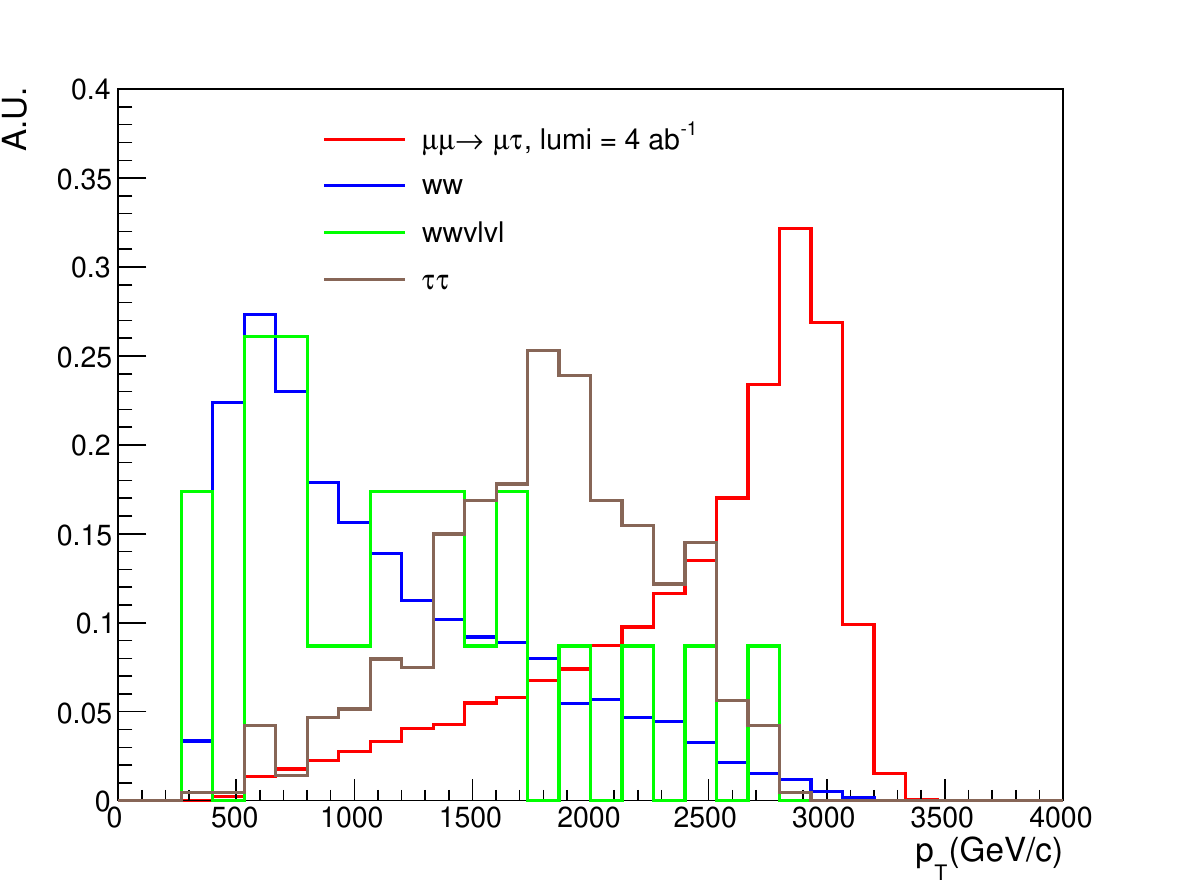}
    \caption{}
  \end{subfigure}
  \caption{$p_T$ distributions of the final state di-leptons for $ee\rightarrow e\mu$ (a), $ee\rightarrow e\tau$ (c), and $ee\rightarrow \mu\tau$ (e) channel at CEPC, and $\mu\mu\rightarrow e\mu$ (b), $\mu\mu\rightarrow e\tau$ (d) and $\mu\mu\rightarrow \mu\tau$ (f) channel at the $6$ $\mathrm{TeV}$ Muon Collider. A.U. refers to the Arbitrary Unit.}
  \label{ptfig}
\end{figure}

Moreover, we show the invariant mass distributions of final state di-leptons for different channels in Figure~\ref{massfig}. To separate the signal from the backgrounds, as  the filtering condition, the invariant masses cut is imposed at the value maximizing the quantity $S/\sqrt{S + B}$, where $S$ denotes the event number of the signal and $B$ denotes the event number of the background. Specifically, the invariant mass cuts of the final state di-leptons for the $ee\rightarrow e\mu$ (Figure~\ref{eeemumll}), $ee\rightarrow e\tau$ (Figure~\ref{eeetaumll}) and $ee\rightarrow \mu\tau$ (Figure~\ref{mumuetaumll}) channel at CEPC, $\mu\mu\rightarrow e\mu$ (Figure~\ref{mumuemumll}), $\mu\mu\rightarrow e\tau$ (Figure~\ref{mumuetaumll}) and $\mu\mu\rightarrow \mu\tau$ (Figure~\ref{mumumutaumll}) channel at the $6$ $\mathrm{TeV}$ Muon Collider, as well as the case at the $14$ $\mathrm{TeV}$ Muon Collider, are selected at $220$ $\mathrm{GeV}$, $160$ $\mathrm{GeV}$, $160$ $\mathrm{GeV}$ for CEPC, $5.2$ $\mathrm{TeV}$, $4$ $\mathrm{TeV}$, $4.2$ $\mathrm{TeV}$ for the $6$ $\mathrm{TeV}$ Muon Collider, and $10$ $\mathrm{TeV}$, $9.5$ $\mathrm{TeV}$, $9.5$ $\mathrm{TeV}$ for the $14$ $\mathrm{TeV}$ Muon Collider, respectively, in order to maximize signal sensitivities from the backgrounds. After all cuts, we obtain histograms on the final state di-leptons $p_T$ distributions for different channels shown in Figure~\ref{ptfig}, which can be exploited to set the upper limits on RPV coupling with the method described below.

\unskip

For each process $X$, we define a per-event weight $n_{L_X}=\sigma_X L / N_X$ to take into account the cross-section difference between the signal and background processes, where $L$ denotes the integrated luminosity of the collider, $\sigma_X$ denotes the cross-section of process $X$, and $N_X$ denotes the number of generated events for process $X$. In our study, we simulate $N_X=10^5$ events for each signal and background process. The signal and backgrounds yields are re-weighted according to their cross-section to be matched.

The test statistic $Z$ is defined as follows,
\begin{equation}
\begin{aligned}
& Z=\sum_{i=1}^{\text {nbins }} Z_i, \\
& \begin{cases}Z_i:=2\left[n_i-b_i+b_i \ln \left(b_i / n_i\right)\right] & 95 \% \text { C.L. Exclusion } \\
Z_i:=2\left[b_i-n_i+n_i \ln \left(n_i / b_i\right)\right] & 5 \sigma \text { Discovery. }\end{cases}
\end{aligned}
\label{StatisEq}
\end{equation}
where $b$ denotes the SM background yields, $n=s+b$ denotes the total yields including both signal and background, and $s$ denotes the CLFV signal yields.
In both cases, each $Z_i$ is subjected to a $\chi^2$ distribution with 1 degree of freedom by Wald's theorem~\cite{Wald1943TestsOS}, and thus, test statistic $Z$ is subjected to $\chi^2$ distribution with the number of degrees of freedom corresponding to the number of bins~\cite{Cowan:2010js}. By summing   each $Z_i$ in $p_T$ distribution histograms and finding the integral of $\chi^2$ probability density function to match the corresponding significance, we can obtain the upper limits on RPV couplings. The results are summarized and plotted in the next section.

\section{Results}
\label{sec:result}

\subsection{CEPC}

\begin{figure}[ht]
  \centering
  \begin{subfigure}[htbp]{0.49\textwidth}
    \centering
    \includegraphics[width=\textwidth]{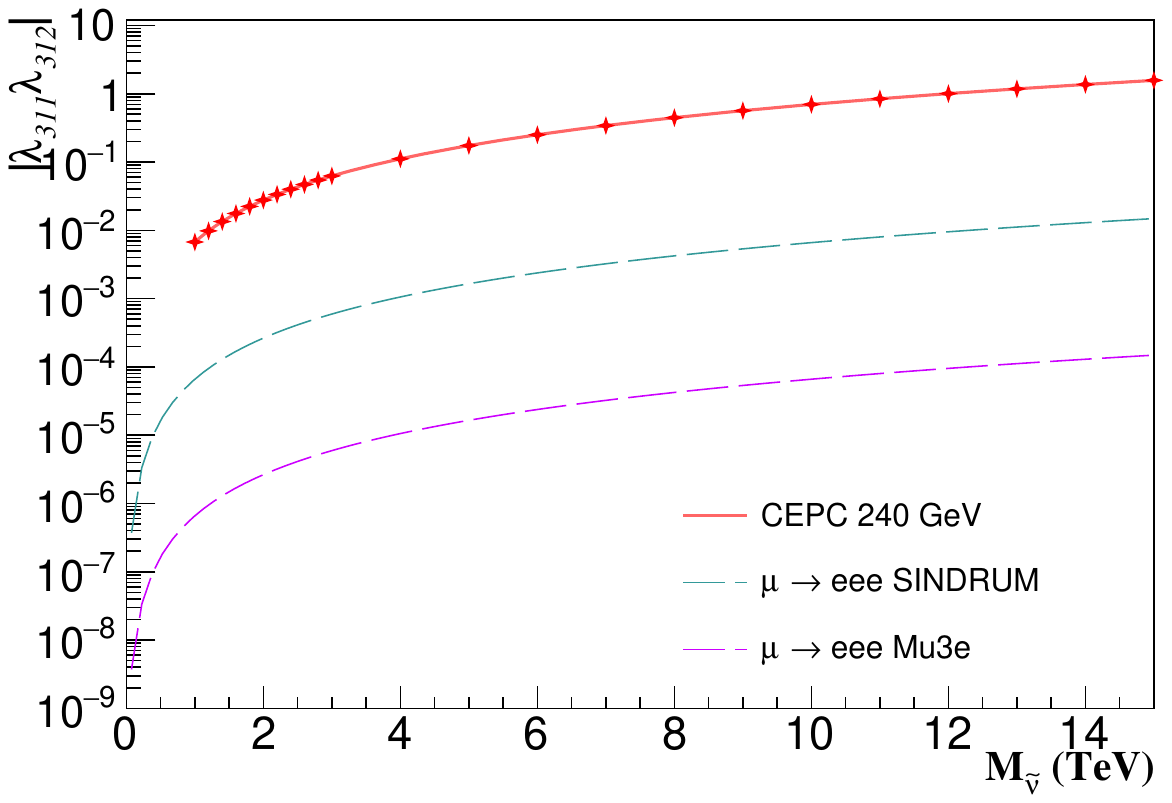}
    \caption{}
  \end{subfigure}
  \hfill
  \begin{subfigure}[htbp]{0.49\textwidth}
    \centering
    \includegraphics[width=\textwidth]{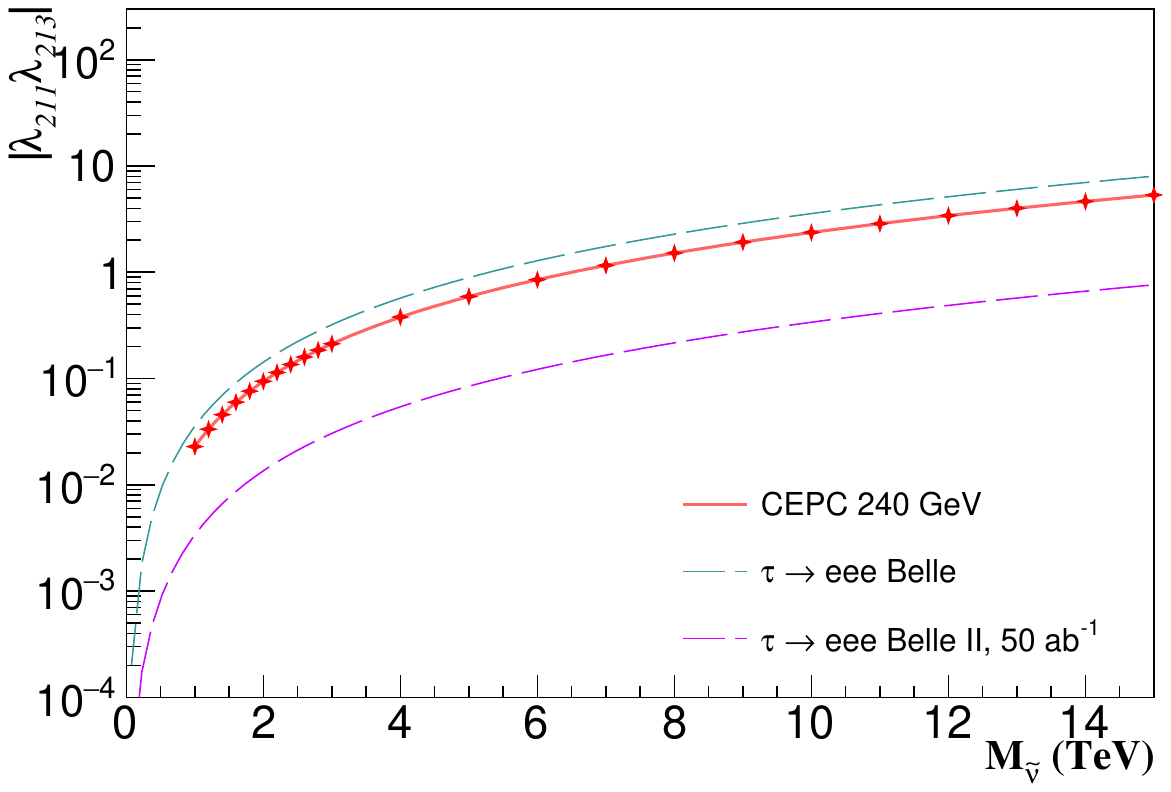}
    \caption{}
  \end{subfigure}
  \newline
  \begin{subfigure}[htbp]{0.49\textwidth}
    \centering
    \includegraphics[width=\textwidth]{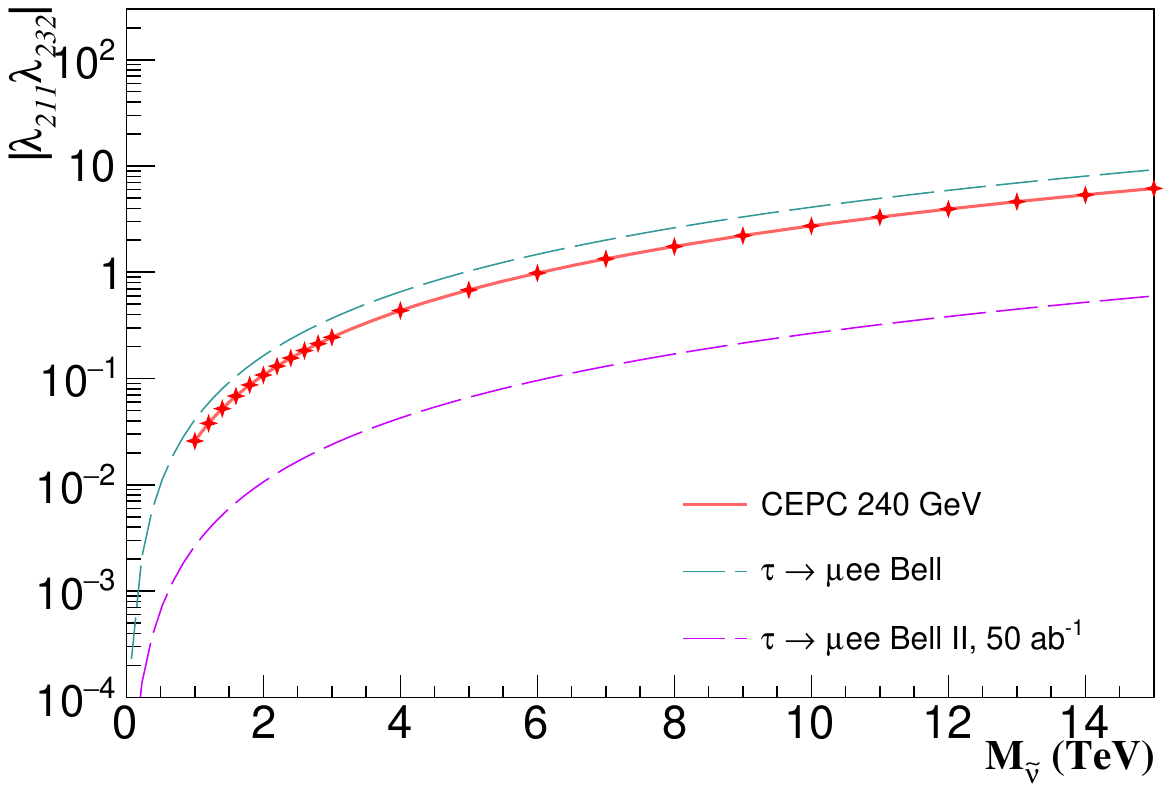}
    \caption{}
  \end{subfigure}
  \caption{The 95\% C.L. upper limit of RPV couplings $\left|\lambda_{311} \lambda_{312}\right|$ (equivalent to $\left|\lambda_{311} \lambda_{321}\right|$ and $\left|\lambda_{211} \lambda_{212}\right|$) from $ee\rightarrow e\mu$ simulation (a), $\left|\lambda_{211} \lambda_{213}\right|$ (equivalent to $\left|\lambda_{211} \lambda_{231}\right|$ and $\left|\lambda_{311} \lambda_{313}\right|$) from $ee\rightarrow e\tau$ simulation (b), $\left|\lambda_{211} \lambda_{232}\right|$ (equivalent to $\left|\lambda_{311} \lambda_{323}\right|$) from $ee\rightarrow \mu\tau$ simulation (c) versus different s-neutrino masses at CEPC.} 
  \label{CEPClimit}
\end{figure}

Using the method described in Section~\ref{sec:method}, the 95\% confidence level (C.L.) upper limit results of various couplings versus different s-neutrino masses obtained from the simulation at CEPC are presented in Figure~\ref{CEPClimit}. The current most stringent upper limit of these couplings  from low-energy $\mu$ and $\tau$ experiments are also included for comparison. For a heavy lepton $a$ decaying into leptons $b$ and $c$ and an anti–lepton $\bar{d}$~\cite{Dreiner:2006gu},
\begin{equation}
\begin{cases}\Gamma_{a \rightarrow b c \bar{d}}=\dfrac{m_{l^a}^5}{6144 \pi^3 m_{\tilde{\nu} g}^4}\left(\lambda_{g d c}^2 \lambda_{g b a}^2+\lambda_{g c d}^2 \lambda_{g a b}^2+\lambda_{g d b}^2 \lambda_{g c a}^2+\lambda_{g b d}^2 \lambda_{g a c}^2\right) & b\neq c \\ \Gamma_{a \rightarrow b b \bar{d}}=\dfrac{m_{l_a}^5}{6144 \pi^3 m_{\tilde{\nu}_{g L}}^4}\left(\lambda_{g d b}^2 \lambda_{g b a}^2+\lambda_{g b d}^2 \lambda_{g a b}^2\right) . & b=c\end{cases}
\label{eqtrans}
\end{equation}
by which we can convert the upper limits of the branching ratio obtained from the experiment to the upper limits of RPV couplings.

As shown in Figure~\ref{CEPClimit}, although the $ee\rightarrow e\mu$ channel simulation gives a looser upper limit of couplings $\left|\lambda_{311} \lambda_{312}\right|$, $\left|\lambda_{311} \lambda_{321}\right|$ and $\left|\lambda_{211} \lambda_{212}\right|$ than the result obtained from the SINDRUM experiment~\cite{SINDRUM:1987nra} and prospect Mu3e experiment, the $ee\rightarrow e\tau$ and $ee \rightarrow \mu\tau$ channel simulation can give more stringent upper limits of couplings $\left|\lambda_{211} \lambda_{213}\right|$, $\left|\lambda_{211} \lambda_{231}\right|$, $\left|\lambda_{311} \lambda_{313}\right|$ and $\left|\lambda_{211} \lambda_{232}\right|$, $\left|\lambda_{311} \lambda_{323}\right|$ than the current best results from the Belle experiment~\cite{Hayasaka:2010np}. However, with future experiments included, the prospective Belle II experiments will give more stringent upper limits of couplings than the $ee\rightarrow e\tau$ and $ee \rightarrow \mu\tau$ channels in our simulation results at CEPC.

\subsection{Muon Collider}

\begin{figure}[ht]
  \centering
  \begin{subfigure}[htbp]{0.49\textwidth}
    \centering
    \includegraphics[width=\textwidth]{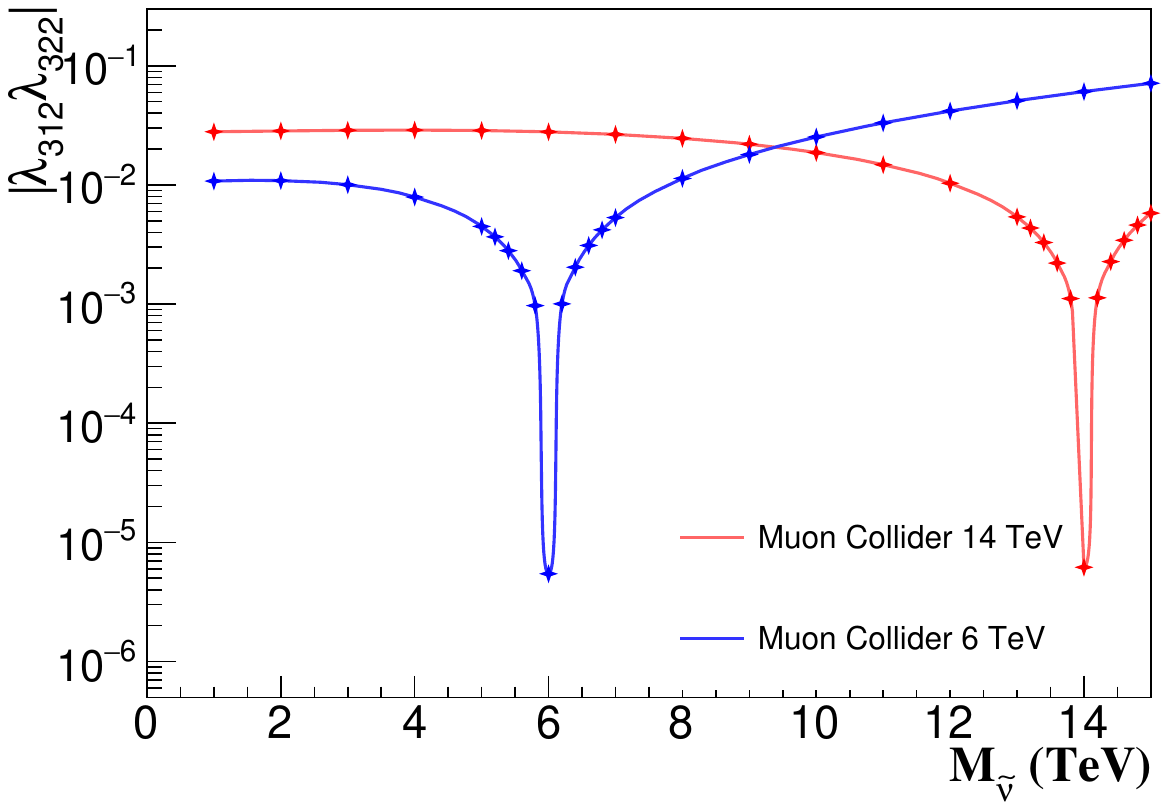}
    \caption{}
  \end{subfigure}
  \hfill
  \begin{subfigure}[htbp]{0.49\textwidth}
    \centering
    \includegraphics[width=\textwidth]{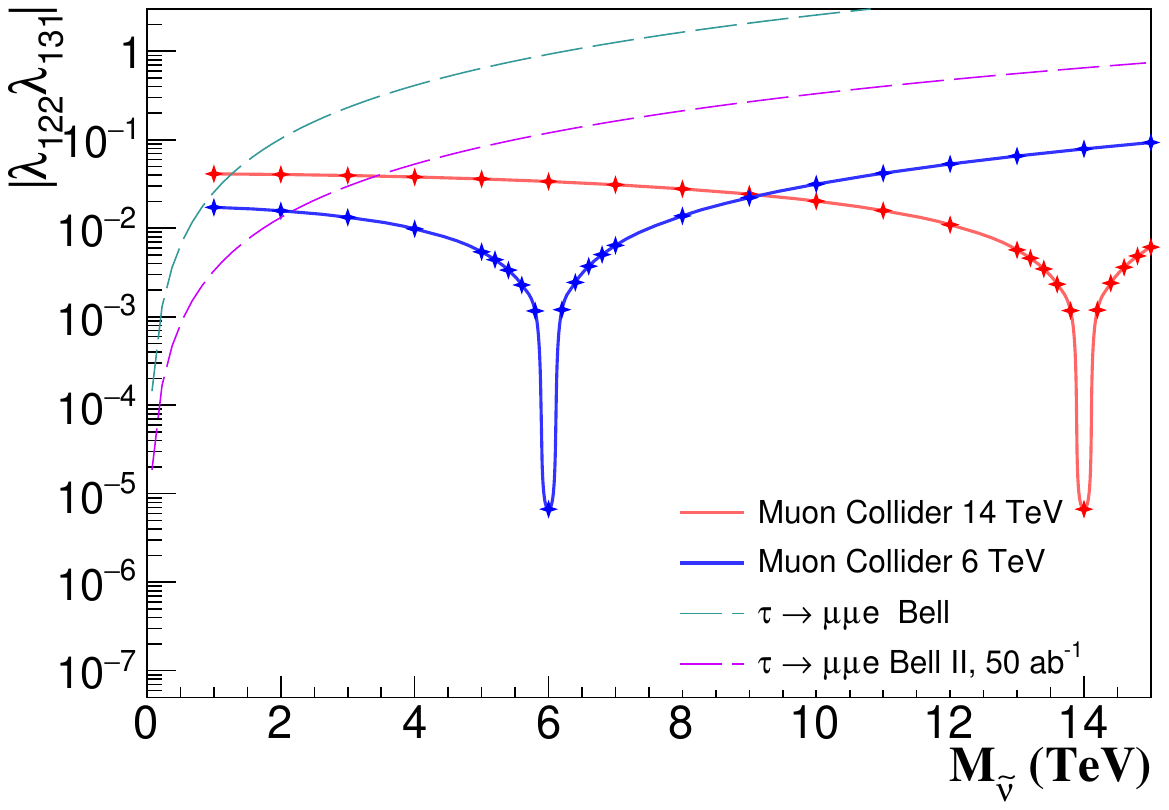}
    \caption{}
  \end{subfigure}
  \newline
  \begin{subfigure}[htbp]{0.49\textwidth}
    \centering
    \includegraphics[width=\textwidth]{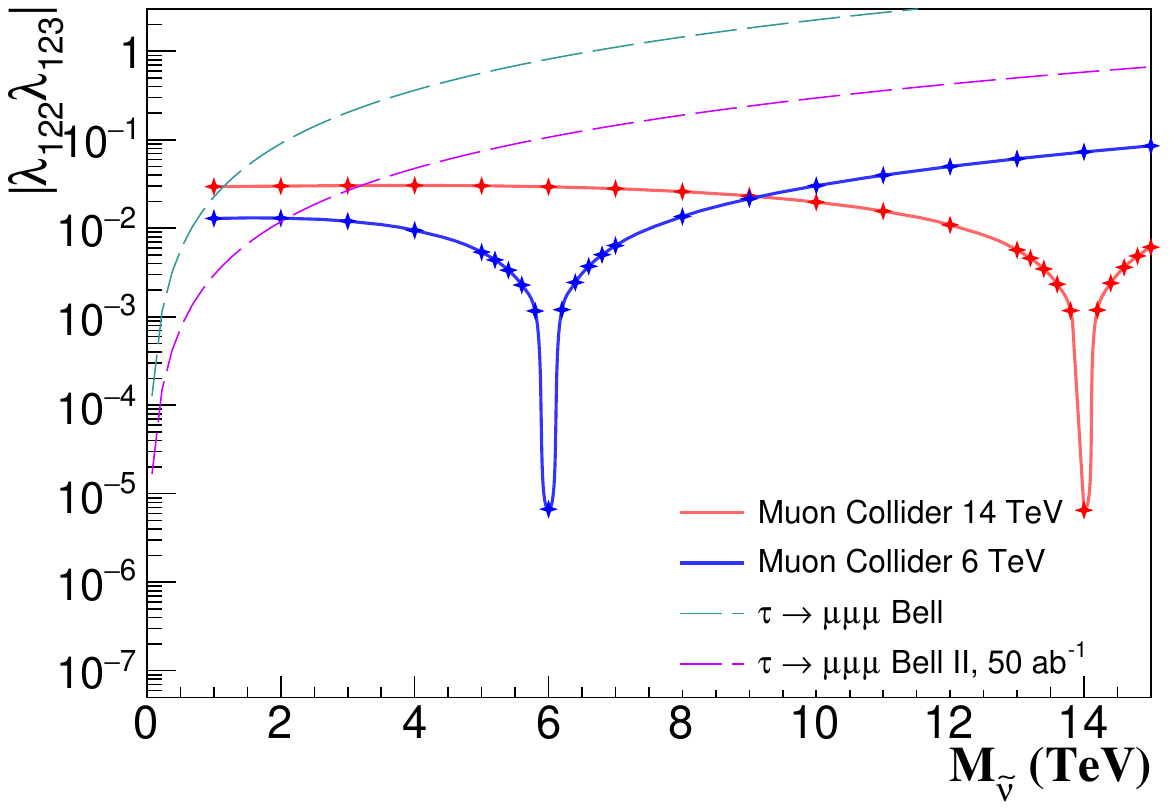}
    \caption{}
  \end{subfigure}
  \caption{95\% C.L. upper limit of RPV couplings $\left|\lambda_{312} \lambda_{322}\right|$ (equivalent to $\left|\lambda_{321} \lambda_{322}\right|$ and $\left|\lambda_{121} \lambda_{122}\right|$) from $\mu\mu\rightarrow e\mu$ simulation (a), $\left|\lambda_{122} \lambda_{131}\right|$ (equivalent to $\left|\lambda_{322} \lambda_{313}\right|$) from $\mu\mu\rightarrow e\tau$ simulation (b), $\left|\lambda_{122} \lambda_{123}\right|$ (equivalent to $\left|\lambda_{122} \lambda_{132}\right|$ and $\left|\lambda_{322} \lambda_{323}\right|$) from $\mu\mu\rightarrow \mu\tau$ simulation (c) versus different s-neutrino masses at the Muon Collider.} 
  \label{MuClimit}
\end{figure}

In the same manner, the 95\% confidence level (C.L.) upper limit result of various couplings versus different s-neutrino masses obtained from simulation at the Muon Collider are presented in Figure~\ref{MuClimit}, with the current most stringent upper limit of these couplings available  for comparison. $\mu\mu\rightarrow e\mu$ simulation can set new limits on couplings $|\lambda_{312}\lambda_{322}|$, $\left|\lambda_{321} \lambda_{322}\right|$ and $\left|\lambda_{121} \lambda_{122}\right|$, which have never been obtained from experiments yet since the process of $\mu \rightarrow \bar{\mu}\mu e $ cannot occur from $\mu$ decay. 

The simulation of the $\mu\mu\rightarrow e\tau$ and $\mu\mu \rightarrow \mu\tau$ channel gives more stringent upper limits of couplings $\left|\lambda_{122} \lambda_{131}\right|$, $\left|\lambda_{322} \lambda_{313}\right|$ and $\left|\lambda_{122} \lambda_{123}\right|$, $\left|\lambda_{122} \lambda_{132}\right|$, $\left|\lambda_{322} \lambda_{323}\right|$ than the current best results from the Belle experiment, and these results are even better than the prospect constraints from the Belle II experiment when the mass of s-neutrino is greater than about $2$~$\mathrm{TeV}$.

\section{Conclusions and Outlook}
\label{sec:conclusion}
In this work, we discuss the sensitivity and the potential for searching for CLFV at future lepton colliders based on RPV-MSSM. By performing fast Monte Carlo simulation of the process $ee\rightarrow e\mu$, $ee\rightarrow e\tau$, $ee\rightarrow \mu\tau$ at $240$ $\mathrm{GeV}$ CEPC, and $\mu\mu\rightarrow e\mu$, $\mu\mu\rightarrow e\tau$, $\mu\mu\rightarrow \mu\tau$ at the $6$ or $14$ $\mathrm{TeV}$ Muon Collider with MadGraph5, $\textsc{Pythia8}$ and $\textsc{Delphes}$, the 95\% C.L. upper limits of RPV couplings versus different s-neutrino masses can be obtained.
We found that $\mu\mu\rightarrow e\mu$ simulation can set new limits on couplings $|\lambda_{312}\lambda_{322}|$, $\left|\lambda_{321} \lambda_{322}\right|$ and $\left|\lambda_{121} \lambda_{122}\right|$, which have not yet been obtained by experiments. The $ee\rightarrow e\tau$ and $ee \rightarrow \mu\tau$, $\mu\mu\rightarrow e\tau$ and $\mu\mu \rightarrow \mu\tau$ channel simulation can give more stringent upper limits than the current best results from the Belle experiment. The $\mu\mu\rightarrow e\tau$ and $\mu\mu \rightarrow \mu\tau$ channel simulation results are even better than the results of the prospect Belle II experiment.
These simulation results demonstrate that   future lepton colliders have some superiority over the current collider experiments and low-energy $\mu$ and $\tau$ experiments in the search for CLFV.

%
%
\bibliography{biblio.bib}

\end{document}